\documentstyle[12pt,epsfig]{article}

\def\beq{\begin{equation}}
\def\eeq{\end{equation}}
\def\bea{\begin{eqnarray}}
\def\eea{\end{eqnarray}}
\def\bq{\begin{quote}}
\def\eq{\end{quote}}

\def\bq{\begin{quote}}
\def\eq{\end{quote}}

\begin{document}

\title{\hspace{4.1in}{\small SHEP 03-14; CERN-TH/2003-147\bigskip }\\
\hspace{4.1in}\\
Fermion Masses and Mixing Angles from $SU(3)$ Family Symmetry and Unification%
}
\author{S. F. King$^{\dagger }$and G. G. Ross$^{\ddagger }$ \\
$^{\dagger }$Department of Physics and Astronomy, University of Southampton,%
\\
Southampton, SO17 1BJ, U.K.\\
$^{\ddagger }$CERN, 1211 Geneva 23, Switzerland\\
and\\
Department of Physics, Theoretical Physics, University of Oxford,\\
1 Keble Road, Oxford OX1 3NP, U.K.}
\date{}
\maketitle

\begin{abstract}
We develop a bottom-up approach to constructing a theory of fermion masses
and mixing angles based on the gauge group $SU(3)\times G$ where $SU(3)$ is
a family symmetry and $G$ contains a unified group such as $SO(10)$ or its
Pati-Salam subgroup, together with other discrete symmetries. We construct a
realistic model and show that it can provide an excellent description of
quark and lepton masses and mixing angles, including almost maximal
atmospheric mixing and the LMA MSW solar neutrino solution. We predict a
neutrino mixing angle $\theta_{13}$ near the current limit. The model
provides the basis for a new solution to the flavour problem with a
characteristic soft SUSY breaking mass spectrum.
\end{abstract}

\vspace{1.0in}

\begin{center}
{\it Dedicated to Ian I. Kogan}
\end{center}


\newpage

\section{Introduction}

The flavour problem, the problem of the origin of families and of fermion
masses and mixing angles, has been a longstanding unanswered question facing
the Standard Model, and remains a powerful motivation for going beyond it 
\cite{Ross:2000fn}. The recent progress in neutrino physics in fact demands
new physics beyond the Standard Model, and implies that any solution to the
flavour problem must also include (almost) maximal atmospheric neutrino
mixing, and large mixing angle (LMA) MSW solar neutrino mixing \cite%
{Alberico:2003kd}. Such a spectrum can be readily reproduced from the
see-saw mechanism in a very natural way using right-handed neutrino
dominance \cite{King:2002nf}, but the necessary conditions required for this
mechanism to work can only be understood in terms of beyond Standard Model
physics. On the other hand, these conditions provide powerful clues to the
nature of the new physics, which may help to unlock the whole mystery of
flavour.

It is clear that any hope of a understanding the flavour problem from
present data is only going to be possible if the Yukawa matrices exhibit a
high degree of symmetry. A recent phenomenological analysis shows that an
excellent fit to all quark data is given by the approximately symmetric form
of quark Yukawa matrices \cite{Roberts:2001zy} 
\begin{equation}
Y^{u}\propto \left( 
\begin{array}{ccc}
0 & \epsilon ^{3} & O(\epsilon ^{3}) \\ 
. & \epsilon ^{2} & O(\epsilon ^{2}) \\ 
. & . & 1%
\end{array}%
\right) ,\ \ \ \ Y^{d}\propto \left( 
\begin{array}{crc}
0 & 1.5\bar{\epsilon}^{3} & 0.4\bar{\epsilon}^{3} \\ 
. & \bar{\epsilon}^{2} & 1.3\bar{\epsilon}^{2} \\ 
. & . & 1%
\end{array}%
\right)  \label{yuk}
\end{equation}%
where the expansion parameters $\epsilon $ and $\bar{\epsilon}$ are given by 
\begin{equation}
\epsilon \approx 0.05,\ \ \bar{\epsilon}\approx 0.15.  \label{exp}
\end{equation}

In \cite{King:2001uz} we showed how Yukawa matrices with the structure of Eq.%
\ref{yuk}, could originate from an $SU(3)$ family symmetry.
\footnote{For reviews of $SU(3)$ family symmetry with original
references see for example \cite{Berezhiani:2001mh}.} 
The $SU(3)$
family symmetry constrains the leading order terms to have equal
coefficients \cite{King:2001uz}. We also showed that, due to the see-saw
mechanism \cite{seesaw}, the neutrino Yukawa matrix $Y^{\nu }$ could have a
similar form to $Y^{u}$ in Eq.\ref{yuk}, providing that the heavy Majorana
matrix $M_{RR}$ has a strongly hierarchical form. In the explicit model
presented \cite{King:2001uz} a new expansion parameter was invoked to
describe the right-handed neutrino sector. The first right-handed neutrino
was arranged to be light enough to be the dominant one, and the second one
the leading subdominant one, corresponding to sequential dominance \cite%
{King:2002nf}. This implies that the atmospheric neutrino mixing angle is
given by $\tan \theta _{23}\approx Y_{21}^{\nu }/Y_{31}^{\nu }$, where this
ratio is equal to unity at leading order due to the $SU(3)$ symmetry.
Similarly the solar neutrino angle is then given by $\tan \theta
_{12}\approx \sqrt{2}Y_{12}^{\nu }/(Y_{22}^{\nu }-Y_{32}^{\nu })$, where the
leading order terms in the denominator cancel due to the $SU(3)$ symmetry.
We further proposed that the charged lepton Yukawa matrix $Y^{e}$ has a
similar form to $Y^{d}$ in Eq.\ref{yuk}, apart from a Georgi-Jarlskog factor
of 3 premultiplying the $\bar{\epsilon}^{2}$ terms \cite{GJ}.

Although the above $SU(3)$ family symmetry model is in many ways very
attractive, the model presented in \cite{King:2001uz} has one major
shortcoming: the proposed form of neutrino Yukawa matrix $Y^{\nu }$ implies
that LMA MSW solar solution cannot be reproduced. In this paper we shall
construct a modified version in which the LMA MSW solution is natural. The
difficulty in obtaining a large solar angle was due to the fact that $%
Y_{(22,32)}^{\nu }\sim \epsilon ^{2}$ are larger than $Y_{12}^{\nu }\sim
\epsilon ^{3}$. In \cite{Ross:2002fb} it was shown that if a Grand Unified
theory (GUT) such as $SO(10)$ \cite{Fritzsch:nn}
is used to obtain the Georgi-Jarlskog factor,
it simultaneously suppresses $Y_{(22,32)}^{\nu }$, permitting the LMA MSW
solution. The basic idea is that the effective Yukawa couplings are
generated by Froggatt-Nielsen diagrams which involve a Higgs field $\Sigma $
in the $45$ of $SO(10)$ coupling the fermion line, as shown in Figure \ref%
{treegraph}. The external lines are then left-handed fermions $\psi $ or
charge conjugates of right-handed fermions $\psi ^{c}$ belonging to the
second or third family, the internal lines are corresponding ``fermion
messengers'', and $H$ in the $10$ of $SO(10)$ contains the usual Higgs
doublets. The left-handed fermion messengers $\chi ,\bar{\chi}$ have a mass $%
M$, while the charge conjugates of the right-handed messengers $\chi ^{c},%
\bar{\chi ^{c}}$ have mass $M^{\prime }$. If $M^{\prime }\ll M$, due to
left-right $SO(10)$ breaking, then the second diagram (b) is expected to
dominate. If, in addition, $\Sigma $ gets a vacuum expectation value (vev)
in the hypercharge direction $Y$, so that its couplings to fermions are
proportional to their hypercharge\footnote{%
In \cite{Ross:2002fb} this was characterized by $B-L+2T_{3R}$ ; this is
proportional to hypercharge.}, then this results in the usual
Georgi-Jarlskog factor of 3, since right-handed charged leptons have 3 times
the charge and hypercharge of right-handed down quarks. In addition it leads
to a suppressed coupling in the neutrino Yukawa matrix, since right-handed
neutrinos have zero charge and hypercharge. This suppression then permits
the LMA MSW solution. 
\begin{figure}[th]
\label{treegraph} \centering                     
\mbox{\epsfig{file=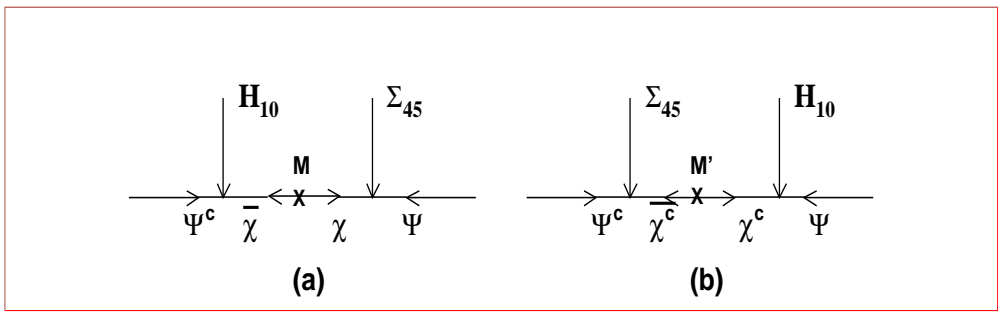, height=1.6in,
width=5.2in}}
\caption{Froggatt Nielsen supergraphs generating fermion masses (from 
\protect\cite{Ross:2002fb}).}
\end{figure}

$SO(10)$ unification has other important implications for the theory. It
reduces the maximum possible family symmetry from $U(3)^{6}$ to $U(3)$, and
implies that $\psi $ and $\psi ^{c}$ must both transform as triplets under
the family symmetry which we shall take to be the gauged $SU(3)$ subgroup of 
$U(3)$. This immediately implies that all fermion masses vanish in the limit
of unbroken $SU(3)$. However in the context of supersymmetry the presence of 
$SU(3)$ can be seen as desirable since it helps to ensure that the sfermion
masses are approximately degenerate as required by flavour changing neutral
current phenomenology. Another important implication of $SO(10)$ is related
to its spontaneous breaking. We shall assume that it is broken, at least
partially, by Wilson line breaking which corresponds to a higher dimensional
component of a higher dimensional gauge field developing a vev. The
advantage of Wilson line breaking is that since gauge fields couple
universally it generates a universal mass for states in a given
representation. For instance if it generates the dominant breaking of $%
SU(2)_{R}$ it leads to three universal messenger mass scales: the
right-handed up masses $M^{u}$, the right-handed down masses $M^{d}$, and
the left-handed doublet masses $M_{L}$. This observation greatly increases
the predictive power of the Froggatt-Nielsen approach since the messengers
in a given representation have universal masses of order the
compactification scale. This motivates the idea that the physics of flavour
resides at the compactification scale, presumably not too far below the
string scale, with Wilson line symmetry breaking playing an important r\^{o}%
le.

In this paper we shall develop a bottom-up approach to constructing a theory
of fermion masses and mixing angles based on the family unification gauge
group $SU(3)\times G$. The bottom-up approach means that we shall work just
below the string scale, where $G=SO(10)$ has been broken either in the $%
SU(5) $ or the Pati-Salam direction \cite{Pati:1974yy}
\footnote
{In this paper we emphasise the possibilty that the underlying theory has an
underlying $G=SO(10)$ symmetry. However the approach is readily adapted to
other underlying grand unified schemes.}. In practice we shall start from
the Pati-Salam subgroup which avoids the problems of doublet-triplet
splitting \cite{King:fv}. \footnote{%
If the symmetry is broken by Wilson lines the doublet-triplet splitting
problem can be elegantly solved \cite{Witten:2001bf}.}, but still allows us
to exhibit the effects associated with left-right, $SU(2)_{R}$, and
quark-lepton symmetry breaking, which mainly concern us here. The underlying
unification implies several important differences to the approach followed
in \cite{King:2001uz}. In particular a new symmetry is required to enforce
the presence of the $\Sigma $ field in the second and third family
operators. The new symmetry should also allows us to predict $M_{RR}$ in
terms of the same expansion parameters that determines the Yukawa matrices,
and not rely on a new expansion parameter.

The layout of the remainder of the paper is as follows. In Section 2 we
discuss our general approach involving Family Symmetry and Unification, and
introduce the basic framework of our model. In Section 3 we discuss a
realistic model in some detail, specify the operators allowed by the
symmetries, and discuss the Yukawa and Majorana matrices which result from a
particular messenger sector. In Section 4 we justify the spontaneous
symmetry breaking of the symmetries assumed in the model. In Section 5 we
consider some supersymmetric aspects theory, in particular for soft masses
and the r\^{o}le of D-terms. In section 6 we discuss the phenomenology of
the model, including the neutrino masses and mixing angles. Section 7
concludes the paper.

\section{Family Symmetry and Unification\label{ssb}}

In this section we shall introduce the ideas of family symmetry and
unification including the necessary ingredients needed to solve the flavour
problem. We shall construct a supersymmetric version of the theory but for
simplicity of presentation, we will treat the supersymmetric structure as
implicit.

In a theory of family symmetry and unification based on gauged $SU(3)\times
SO(10) $ all quarks and leptons originate from a single representation $\psi
_{i}\sim ({\bf 3,16})$. The Higgs doublets are contained in the $H\sim ({\bf %
1,10})$, while $\Sigma \sim ({\bf 1,45})$. In our bottom-up approach we do
not construct the fully unified Grand Unified or string model but start with
the models defined slightly below the GUT or string scale where the
surviving maximal subgroups of $SO(10)$ may be either $SU(5)\times U(1)_{X}$%
, or the Pati-Salam subgroup $G_{PS}=SU(4)_{PS}\times SU(2)_{L}\times
SU(2)_{R}$. In this paper we shall focus on the Pati-Salam symmetry breaking
direction which is the most predictive case. Pati-Salam breaking down to the
Standard Model gauge group is achieved by a combination of Wilson line
breaking, which is dominantly responsible for $SU(2)_{R}$ breaking, and
Higgs breaking due to a field $\Sigma $ which is an adjoint of both $%
SU(4)_{PS}$ and $SU(2)_{R}$, and will play the same r\^{ole} as the adjoint $%
45$ of $SO(10)$ discussed above, from which it may originate. The full
flavour symmetry of the model must involve some additional symmetry under
which $\Sigma $ transforms, and is responsible for $\Sigma $ appearing in
the leading operators in the 22,23,32 positions of the Yukawa matrices. The
symmetry must allow all the necessary leading operators, while suppressing
all unwanted operators, including all subleading operators which are not
required. It must also lead to an acceptable heavy Majorana matrix $M_{RR}$.

\bigskip {\small 
\begin{table}[tbp] \centering%
\begin{tabular}{|ccccccrr|rrc|}
\hline
${\bf Field}$ & ${\bf SU(3)}$ & ${\bf SU(4)_{PS}}$ & ${\bf SU(2)_{L}}$ & $%
{\bf SU(2)_{R}}$ & ${\bf R}$ & ${\bf Z}_{2}$ & ${\bf U(1)}$ & ${\bf Z}_{5}$
& ${\bf Z}_{3}$ & ${\bf Z}_{2}^{\prime }$ \\ \hline
$\psi $ & ${\bf 3}$ & ${\bf 4}$ & ${\bf 2}$ & ${\bf 1}$ & ${\bf 1}$ & 
\multicolumn{1}{c}{${\bf +}$} & ${\bf 0}$ & ${\bf 0}$ & ${\bf 0}$ & ${\bf +}$
\\ 
$\psi ^{c}$ & ${\bf 3}$ & ${\bf \overline{4}}$ & ${\bf 1}$ & ${\bf 2}$ & $%
{\bf 1}$ & \multicolumn{1}{c}{${\bf +}$} & ${\bf 0}$ & ${\bf 0}$ & ${\bf 0}$
& ${\bf +}$ \\ 
$\theta $ & ${\bf \overline{3}}$ & ${\bf 4}$ & ${\bf 1}$ & ${\bf 2}$ & ${\bf %
0}$ & \multicolumn{1}{c}{${\bf +}$} & ${\bf 0}$ & ${\bf 0}$ & ${\bf 0}$ & $%
{\bf +}$ \\ 
$\overline{\theta }$ & ${\bf 3}$ & ${\bf \overline{4}}$ & ${\bf 1}$ & ${\bf 2%
}$ & ${\bf 0}$ & \multicolumn{1}{c}{${\bf +}$} & $-{\bf 6}$ & $-{\bf 1}$ & $%
{\bf 0}$ & ${\bf +}$ \\ 
$H$ & ${\bf 1}$ & ${\bf 1}$ & ${\bf 2}$ & ${\bf 2}$ & ${\bf 0}$ & 
\multicolumn{1}{c}{${\bf +}$} & ${\bf 8}$ & ${\bf -2}$ & ${\bf -1}$ & ${\bf +%
}$ \\ 
$\Sigma $ & ${\bf 1}$ & ${\bf 15}$ & ${\bf 1}$ & ${\bf 3}$ & ${\bf 0}$ & 
\multicolumn{1}{c}{${\bf +}$} & ${\bf 2}$ & ${\bf 2}$ & ${\bf -1}$ & ${\bf +}
$ \\ 
$\phi _{3}$ & ${\bf \overline{3}}$ & ${\bf 1}$ & ${\bf 1}$ & ${\bf 3\oplus 1}
$ & ${\bf 0}$ & \multicolumn{1}{c}{${\bf -}$} & ${\bf -4}$ & ${\bf 1}$ & $%
{\bf -1}$ & ${\bf +}$ \\ 
$\phi _{23}$ & ${\bf \overline{3}}$ & ${\bf 1}$ & ${\bf 1}$ & ${\bf 1}$ & $%
{\bf 0}$ & \multicolumn{1}{c}{${\bf +}$} & ${\bf -5}$ & ${\bf 0}$ & ${\bf 1}$
& ${\bf -}$ \\ 
$\overline{\phi _{3}}$ & ${\bf 3}$ & ${\bf 1}$ & ${\bf 1}$ & ${\bf 3\oplus 1}
$ & ${\bf 0}$ & \multicolumn{1}{c}{${\bf -}$} & ${\bf -2}$ & ${\bf -2}$ & $%
{\bf 1}$ & ${\bf +}$ \\ 
$\overline{\phi _{23}}$ & ${\bf 3}$ & ${\bf 1}$ & ${\bf 1}$ & ${\bf 1}$ & $%
{\bf 0}$ & \multicolumn{1}{c}{${\bf +}$} & ${\bf 6}$ & ${\bf 1}$ & ${\bf 0}$
& ${\bf +}$ \\ 
$\overline{\phi _{2}}$ & ${\bf 3}$ & ${\bf 1}$ & ${\bf 1}$ & ${\bf 3\oplus 1}
$ & ${\bf 0}$ & \multicolumn{1}{c}{${\bf -}$} & ${\bf 5}$ & ${\bf 0}$ & $%
{\bf -1}$ & ${\bf -}$ \\ \hline
\end{tabular}%
\caption{{\footnotesize Transformation of the superfields under the
$SU(3)$ family, Pati-Salam and $R\times Z_2 \times U(1)$ symmetries which
restrict the form of the mass matrices for three representative
examples. The continuous R-symmetry may be alternatively be replaced
by a discrete $Z_{2R}$ symmetry. Also shown in the last three columns 
is the transformation under a $Z_5\times Z_3\times Z_2'$
subgroup of the $U(1)$ which is sufficient to ensure a
phenomenologically viable pattern of couplings. We only display the
fields relevant for generating fermion mass and spontaneous symmetry
breaking.  \label{Table1}}} 
\end{table}%
}

The explicit bottom-up models we shall construct are based on $SU(3)$ family
symmetry commuting with Pati-Salam symmetry, $SU(3)\times G_{PS}$. The
transformation properties of the left-handed quarks and leptons $\psi _{i}$,
the left-handed charge conjugates of the right-handed quarks and leptons $%
\psi _{i}^{c}$, the Higgs doublets $H$ and the $\Sigma $ field under the
gauge group $SU(3)\times G_{PS}$ are given in Table \ref{Table1}. Assuming
the Pati-Salam symmetry to start with has the advantage that it explicitly
exhibits $SU(4)_{PS}$ quark-lepton and $SU(2)_{R}$ isospin symmetry,
allowing Georgi-Jarlskog factors to be generated and isospin breaking to be
controlled, while avoiding the Higgs doublet-triplet splitting problem \cite%
{King:fv}. The $SU(4)_{PS}$ symmetry also provides a welcome restriction of
the messenger masses, providing a link between the up-quarks and neutrinos.
The fields $\theta $ and $\overline{\theta }$ carry lepton number $1$ and $%
-1 $ respectively. They acquire vevs and break lepton number giving rise to
the Majorana masses for the neutrino components of $\psi ^{c}.$

The adjoint $\Sigma$ field develops vevs in the $SU(4)_{PS}\times SU(2)_R$
direction which preserves the hypercharge generator $Y=T_{3R}+(B-L)/2$, and
implies that any coupling of the $\Sigma$ to a fermion and a messenger such
as $\Sigma^{a \alpha}_{b \beta}\psi^c_{a\alpha}\chi^{b\beta}$, where the $%
SU(2)_R$ and $SU(4)_{PS}$ indices have been displayed explicitly, is
proportional to the hypercharge $Y$ of the particular fermion component of $%
\psi^c$ times the vev $\sigma$.

To build a viable model we also need spontaneous breaking of the family
symmetry 
\begin{equation}
SU(3)\longrightarrow SU(2)\longrightarrow {\rm Nothing}  \label{fsb}
\end{equation}%
To achieve this symmetry breaking we introduce additional Higgs fields $\phi
_{3},$ $\overline{\phi }_{3},$ $\phi _{23}$ and $\overline{\phi }_{23}$ in
the representations given in Table \ref{Table1}. The largeness of the third
family fermion masses implies that $SU(3)$ must be strongly broken by new
Higgs antitriplet fields $\phi _{3}$ which develop a vev in the third $SU(3)$
component $<\phi _{3}>^{T}=(0,0,a_{3})$ as in \cite{King:2001uz}. However,
for reasons discussed later, we assume that $\phi _{3}^{i}$ transforms under 
$SU(2)_{R}$ as ${\bf 3\oplus 1}$ rather than being $SU(2)_{R}$ singlets as
we assumed in \cite{King:2001uz}, and develops vevs in the $SU(3)\times
SU(2)_{R}$ directions 
\begin{equation}
<\phi _{3}>=<\overline{\phi _{3}}>=\left( 
\begin{array}{c}
0 \\ 
0 \\ 
1%
\end{array}%
\right) \otimes \left( 
\begin{array}{cc}
a_{3}^{u} & 0 \\ 
0 & a_{3}^{d}%
\end{array}%
\right) .  \label{phi3vev}
\end{equation}%
The symmetry breaking also involves the $SU(3)$ antitriplets $\phi _{23}$
which develop vevs \cite{King:2001uz} 
\begin{equation}
<\phi _{23}>=\left( 
\begin{array}{c}
0 \\ 
1 \\ 
e^{i\theta }%
\end{array}%
\right) b,  \label{phi23vevs}
\end{equation}%
where, as in \cite{King:2001uz}, vacuum alignment ensures that the vevs are
aligned in the 23 direction. Due to D-flatness there must also be
accompanying Higgs triplets such as $\overline{\phi _{23}}$ which develop
vevs \cite{King:2001uz} 
\begin{equation}
<\overline{\phi _{23}}>=\left( 
\begin{array}{c}
0 \\ 
1 \\ 
e^{-i\theta }%
\end{array}%
\right) be^{i\phi }.  \label{phibar23vevs}
\end{equation}%
In Section \ref{spont} we will show how this pattern can be achieved through
the introduction of the additional triplet field $\overline{\phi _{2}}$
given in Table \ref{Table1}. With the spectrum shown in this Table there are
residual $SU(3)$ and $U(1)$ anomalies but no mixed anomalies involving the
Standard Model gauge group. These anomalies can be cancelled by the addition
of Standard Model singlet fields all of which can acquire a mass at the
scale of breaking of $SU(3).$ We do not list these fields here as they play
no role in the low energy theory but note that in a more unified model such
anomaly cancellation can happen in an elegant manner \cite{Ling:2003kr}.

\section{A Realistic Model}

\subsection{Operators and Additional Symmetries}

In building a phenomenologically viable scheme it is necessary to constrain
the allowed Yukawa couplings through additional symmetries. There is
considerable freedom in implementing such symmetries, the resultant models
differing in their detailed phenomenology. In this paper we present a simple
example in which the $SU(3)$ family symmetry is augmented by a $Z_{2}\times
U(1)$ gauge symmetry. It will ensure that the quark and lepton Dirac masses
have an acceptable form and also order the Majorana mass matrix of the right
handed neutrinos so that the see-saw mechanism gives to large mixing angles.
The assignment of the $Z_{2}\times U(1)$ charges is shown in Table \ref%
{Table1}. The symmetries of the model are completed through the addition of
an R-symmetry (or a discrete version of it $Z_{2R}$).

In practice it is not necessary that the full $U(1)$ symmetry be present and
a discrete subgroup can be sufficient to limit the allowed Yukawa couplings.
For example, the discrete group $Z_{5}\times Z_{3}\times Z_{2}^{\prime }$
with charges given in Table \ref{Table1} gives the same leading operators
discussed in the next Section and hence approximately the same mass
matrices. Note that the charges under the discrete symmetry look simpler
than those for the $U(1)$ showing that is not necessary to have an exotic
choice of charges to achieve a realistic model.

The leading operators allowed by the symmetries are 
\begin{eqnarray}
P_{{\rm Yuk}} &\sim &\frac{1}{M^{2}}\psi _{i}\phi _{3}^{i}\psi _{j}^{c}\phi
_{3}^{j}H  \label{op1} \\
&+&\frac{\Sigma }{M^{3}}\psi _{i}\phi _{23}^{i}\psi _{j}^{c}\phi _{23}^{j}H
\label{op2} \\
&+&\frac{1}{M^{5}} \left( (\epsilon ^{ijk}\psi _{i}^{c}\overline{\phi _{23}}%
_{j}\overline{\phi _{3,k}})(\psi _{l}\phi _{23}^{l}) + (\epsilon ^{ijk}\psi
_{i}\overline{\phi _{23}}_{j}\overline{\phi _{3,k}})(\psi^c_{l}\phi
_{23}^{l}) \right) H(\phi _{23}^{m}\overline{\phi }_{3,m})  \label{op3} \\
&+&
\frac{1}{M^{5}}(\epsilon ^{ijk}\psi _{i}^{c}\overline{\phi _{23}}_{j}\psi
_{k})H(\phi _{23}^{l}\overline{\phi }_{3,l})^{2}  
+ 
\frac{1}{M^{5}}(\epsilon ^{ijk}\psi _{i}^{c}\overline{\phi _{3}}_{j}\psi
_{k})H
(\phi _{23}^{l}\overline{\phi }_{23,l})(\phi _{23}^{m}\overline{\phi }_{3,m})
\label{op4} \\
&+&\frac{1}{M^{4}}\left( \psi _{i}\phi _{23}^{i}\psi _{j}^{c}\phi
_{3}^{j}+\psi _{i}\phi _{3}^{i}\psi _{j}^{c}\phi _{23}^{j}\right) H.S
\label{op5} \\
P_{{\rm Maj}} &\sim &\frac{1}{M}\psi _{i}^{c}\theta ^{i}\theta ^{j}\psi
_{j}^{c}  \label{mop1} \\
&+&\frac{1}{M^{11}}\psi _{i}^{c}\phi _{23}^{i}\psi _{j}^{c}\phi
_{23}^{j}(\theta ^{k}\overline{\phi _{23,k}})(\theta ^{l}\overline{\phi
_{3,l}})(\phi _{3}\overline{\phi _{23}})^{3}  \label{mop2} \\
&+&\frac{1}{M^{13}}(\epsilon ^{ijk}\psi _{i}^{c}\overline{\phi _{23,j}}%
\overline{\phi _{3,k}})^{2}(\theta ^{k}\overline{\phi _{23,k}})(\theta ^{l}%
\overline{\phi _{3,l}})(\phi _{3}\overline{\phi _{23}})(\phi _{23}\overline{%
\phi _{3}})^{2}  \label{mop3}
\end{eqnarray}%
where, as discussed below, the operator mass scales, generically denoted by $%
M$ may differ and we have suppressed couplings of $O(1).$ The field $S$ is
involved in symmetry breaking as discussed in Section \ref{spont}. Its
quantum numbers are given in Table \ref{Table2}. It acquires a vev of $%
O(\phi _{23}\overline{\phi _{23}}).$

\subsection{Messengers and the $(2,3)$ Yukawa block}

The leading Yukawa operators which contribute to the $(2,3)$ block of the
Yukawa matrices are given in Eqs.\ref{op1} and \ref{op2}. These operators
arise from Froggat-Nielsen diagrams similar to Figure \ref{treegraph}, but
generalized to include insertions of the $\phi _{3},\phi _{23}$ fields. $M$
represents the right-handed up and down messenger mass scales $M^{u,d}$,
corresponding to the dominance of diagram (b), which applies if $M<M^{L}$
where $M^{L}$ represents the left-handed messenger mass scale. We shall not
specify the messenger sector explicitly, but characterize it by the
messenger mass scales 
\begin{equation}
M^{d}\approx \frac{1}{3}M^{u}\ll M^{L}.  \label{mess}
\end{equation}%
Such a universal structure is to be expected in theories with Wilson line
breaking in which the breaking is due to the $(4D)$ scalar component of a
higher dimension gauge field because it couples universally to fields in the
same representation of gauge group factors left unbroken by the Wilson line.
The Wilson line breaking is associated with the compactification and so the
splitting induced is naturally of order the compactification scale. Thus, if
Wilson line breaking is responsible for breaking $SU(2)_{R},$ the messenger
states (Kaluza-Klein modes or vectorlike states obtaining mass on
compactification) must have masses of order the compactification scale.

Given that $SU(4)_{PS}$ remains after compactification$,$ its subsequent
breaking will be a small effect so that the right-handed lepton messenger
masses are $M^{\nu }\simeq M^{u}$, and $M^{e}\simeq M^{d}$. The splitting of
the messenger mass scales relies on left-right and $SU(2)_{R}$ breaking
effects which we shall assume to be due to the Wilson line symmetry breaking
mechanism. Eq.\ref{mess} implies that diagrams of type (b) in Figure \ref%
{treegraph} dominate, and the expansion parameters associated with $\phi
_{23}$ are then generated as in \cite{King:2001uz} 
\begin{equation}
\epsilon \equiv \frac{b}{M^{u}},\ \ \bar{\epsilon}\equiv \frac{b}{M^{d}}
\label{eps}
\end{equation}

Unlike the previous model \cite{King:2001uz}, we shall construct a model in
which the $\phi _{3}$ vev $a_{3}$ is less than the messenger mass scale $M$.
The reason for this is twofold. Firstly, an underlying $SO(10)$ leads us to
consider fermion messengers, and if $a_{3}>M$ this then implies an
undesirably massive third family fermion $\psi _{3}$ from the coupling $\phi
_{3}\psi _{3}\bar{\chi}$, and a light fermion messenger $\chi $. Secondly,
wavefunction insertions of the invariant operator $\phi _{3}\phi
_{3}^{\dagger }/M^{2}$ on an third family fermion propagator can spoil the
perturbative expansion if $a_{3}>M$. Therefore we shall assume here that $%
a_{3}<M$. If $\phi _{3}$ were a $SU(2)_{R}$ singlet as in \cite{King:2001uz}
then Eq.\ref{mess} would imply that the top quark Yukawa coupling is much
smaller than the bottom quark Yukawa coupling by a factor of $1/9$. This
explains why $\phi _{3}$ cannot be a $SU(2)_{R}$ singlet. For the case that $%
\phi _{3}$ transforms as $2\times \overline{2}$ under $SU(2)_{R}$ it may
acquire vevs $a_{3}^{u}$, $a_{3}^{d}$ in the up and down directions. Then
with $a_{3}^{u}/M^{u}\approx a_{3}^{d}/M^{d}<1$ we have comparable top and
bottom Yukawa couplings, as required. For definiteness we shall consider the
case that 
\begin{equation}
\frac{a_{3}^{u}}{M^{u}}=\frac{a_{3}^{d}}{M^{d}}=\sqrt{\bar{\epsilon}}.
\label{a3}
\end{equation}

It remains to specify the expansion parameter associated with $\sigma $, the
vev of $\Sigma $. For phenomenological reasons we take it to be 
\begin{equation}
\frac{\sigma Y(d)}{M^{d}}=\bar{\epsilon}  \label{sigmad}
\end{equation}%
where $Y(d)=1/3$ is the hypercharge of $d^{c}$. From Eqs.\ref{mess},\ref%
{sigmad}, we find 
\begin{equation}
\frac{\sigma Y(u)}{M^{u}}=-\frac{2}{3}\bar{\epsilon}  \label{sigmau}
\end{equation}%
where $Y(u)=-2/3$ is the hypercharge of $u^{c}$.

The operators in Eqs.\ref{op1}-\ref{mop3} with the expansion parameters in
Eqs.\ref{eps}, \ref{a3}, \ref{sigmad}, \ref{sigmau}, and the vevs in Eqs.\ref%
{phi3vev}, \ref{phi23vevs} lead to the approximate form of the quark Yukawa
matrices for the $(2,3)$ block given by : 
\begin{equation}
Y^{u}\approx \left( 
\begin{array}{ll}
\epsilon ^{2}(-\frac{2}{3}) & \epsilon ^{2}(-\frac{2}{3}) \\ 
\epsilon ^{2}(-\frac{2}{3}) & 1%
\end{array}%
\right) \bar{\epsilon},\ \ \ \ Y^{d}\approx \left( 
\begin{array}{ll}
\bar{\epsilon}^{2} & \bar{\epsilon}^{2} \\ 
\bar{\epsilon}^{2} & 1%
\end{array}%
\right) \bar{\epsilon}  \label{yuk23q}
\end{equation}%
which is of the form in Eq.\ref{yuk}.

The charged lepton Yukawa matrix $Y^{e}$ has a similar form to $Y^{d}$ since
the charged lepton operators are generated by messengers with the quantum
numbers of $e^{c}$, with the same messenger mass scale as for $d^{c}$
messengers, $M^{e}=M^{d}$, due to the $SU(4)_{PS}$ symmetry. However due to $%
\Sigma $ the 22, 23, 32 elements of $Y^{e}$ are multiplied by the
Georgi-Jarlskog factor of $Y(e)/Y(d)=3$, where $Y(e)$ is the hypercharge of $%
e^{c}$, giving 
\begin{equation}
Y^{e}\approx \left( 
\begin{array}{ll}
\bar{\epsilon}^{2}(3) & \bar{\epsilon}^{2}(3) \\ 
\bar{\epsilon}^{2}(3) & 1%
\end{array}%
\right) \bar{\epsilon}.  \label{yuk23e}
\end{equation}

The above results apply in the limit that the Froggatt-Nielsen diagrams are
dominated by the right-handed messengers. In this limit the neutrino Yukawa
matrix $Y^{\nu }$ has zeroes in the 22, 23, 32 positions due to the fact
that these elements are proportional to the hypercharge of the right-handed
neutrino which is zero. This leads to the desired suppression of these
elements of $Y^{\nu }$. To characterise this suppression we define the
expansion parameter in the left-handed neutrino sector 
\begin{equation}
\frac{\sigma Y({\nu _{L}})}{M^{L}}\equiv -\alpha \bar{\epsilon}.
\label{alpha}
\end{equation}%
Then the 22, 23 ,32 elements of $Y^{\nu }$ are of order $\alpha \epsilon
^{2} $, once the overall factor of $\bar{\epsilon}$ has been factored out, 
\begin{equation}
Y^{\nu }\approx \left( 
\begin{array}{ll}
\epsilon ^{2}(-\alpha ) & \epsilon ^{2}(-\alpha ) \\ 
\epsilon ^{2}(-\alpha ) & 1%
\end{array}%
\right) \bar{\epsilon}.  \label{yuk23nu}
\end{equation}

\subsection{The complete Yukawa Matrices}

The leading elements in the 12, 13, 21, 31 positions contain contributions
from two different leading order operators, namely those in Eqs. \ref{op3}
and \ref{op4}. These contributions depend on the vevs of $\phi _{23}$ in Eq.%
\ref{phi23vevs} and $\overline{\phi }_{23}$ in Eq.\ref{phibar23vevs} with
the 12, 13 contributions being of order $\epsilon ^{3}$ in the up and
neutrino sector, and $\bar{\epsilon}^{3}$ in the down and charged lepton
sector, each multiplied by an overall factor of $\bar{\epsilon}$. However,
due to the antisymmetric $SU(3)$ invariant, the relative coupling of the $%
(1,2)$ and $(1,3)$ elements have opposite signs. Note that in a full $SO(10)$
theory the operators in Eq.\ref{op4} are forbidden due to antisymmetry since $%
\psi $ and $\psi ^{c}$ are unified into a single $16$ representation.
However, since $SO(10)$ breaking effects are required in any case, we must
allow for the presence of such operators. The sum of the contributions
gives a factor $g+h+h'$ in the 12 entry and $g-h$ in the 13 entries of $%
Y^{d},Y^{e}$, a factor $g+h/3+h'/3$ in the 12 entry and $g-h/3$ in the 13
entries of $Y^{u},Y^{\nu }$, where we allow for the fact that the $SO(10)$
symmetry breaking effects which are responsible for the existence of the
second term are controlled by the same messenger masses $M^{u},M^{d}$ as in
Eq.\ref{mess}. The corresponding operators in the 21,31 positions
have an independent coefficient $g'$ due to the
contribution from the two operators in Eq. \ref{op3}.

The operators in Eq.\ref{op5} give an important sub-leading contribution to
the 23,32 elements of the Yukawa matrices. This, together with the structure
discussed above, and allowing for the corrections due to wavefunction
insertions of the invariant operator $\phi _{3}\phi _{3}^{\dagger
}/M^{2}\sim \bar{\epsilon}$ on a third family fermion leg, gives the final
form of the Yukawa matrices 
\begin{eqnarray}
Y^{u} &\approx &\left( 
\begin{array}{llr}
0 & \epsilon ^{3}(g+\frac{h}{3}+\frac{h'}{3}) & \epsilon ^{3}(g-\frac{h}{3})(1+O(\bar{%
\epsilon})) \\ 
\epsilon ^{3}(g'-\frac{h}{3}-\frac{h'}{3}) & \epsilon ^{2}(-\frac{2}{3}) & \epsilon ^{2}(-%
\frac{2}{3})+c^{\prime }\epsilon ^{3}\bar{\epsilon}^{-\frac{1}{2}} \\ 
\epsilon ^{3}(g'+\frac{h}{3})(1+O(\bar{\epsilon})) & \epsilon ^{2}(-\frac{2}{3%
})+c\epsilon ^{3}\bar{\epsilon}^{-\frac{1}{2}} & 1+O(\bar{\epsilon})%
\end{array}%
\right) \bar{\epsilon},  \label{Yu} \\
Y^{d} &\approx &\left( 
\begin{array}{llr}
0 & \bar{\epsilon}^{3}(g+h+h') & \bar{\epsilon}^{3}(g-h)(1+O(\bar{\epsilon}))
\\ 
\bar{\epsilon}^{3}(g'-h-h') & \bar{\epsilon}^{2} & \bar{\epsilon}^{2}+c^{\prime }%
\bar{\epsilon}^{\frac{5}{2}} \\ 
\bar{\epsilon}^{3}(g'+h)(1+O(\bar{\epsilon})) & \bar{\epsilon}^{2}+c\bar{%
\epsilon}^{\frac{5}{2}} & 1+O(\bar{\epsilon})%
\end{array}%
\right) \bar{\epsilon},  \label{Yd} \\
Y^{e} &\approx &\left( 
\begin{array}{llr}
0 & \bar{\epsilon}^{3}(g+h+h') & \bar{\epsilon}^{3}(g-h)(1+O(\bar{\epsilon}))
\\ 
\bar{\epsilon}^{3}(g'-h-h') & \bar{\epsilon}^{2}({3}) & \bar{\epsilon}^{2}({3}%
)+c^{\prime }\bar{\epsilon}^{\frac{5}{2}} \\ 
\bar{\epsilon}^{3}(g'+h)(1+O(\bar{\epsilon})) & \bar{\epsilon}^{2}({3})+c\bar{%
\epsilon}^{\frac{5}{2}} & 1+O(\bar{\epsilon})%
\end{array}%
\right) \bar{\epsilon},  \label{Ye} \\
Y^{\nu } &\approx &\left( 
\begin{array}{llr}
0 & \epsilon ^{3}(g+\frac{h}{3}+\frac{h'}{3}) & \epsilon ^{3}(g-\frac{h}{3})(1+O(\bar{%
\epsilon})) \\ 
\epsilon ^{3}(g'-\frac{h}{3}-\frac{h'}{3}) & \epsilon ^{2}(-\alpha ) & \epsilon
^{2}(-\alpha )+c^{\prime }\epsilon ^{3}\bar{\epsilon}^{-\frac{1}{2}} \\ 
\epsilon ^{3}(g'+\frac{h}{3})(1+O(\bar{\epsilon})) & \epsilon ^{2}(-\alpha
)+c\epsilon ^{3}\bar{\epsilon}^{-\frac{1}{2}} & 1+O(\bar{\epsilon})%
\end{array}%
\right) \bar{\epsilon}.  \label{Ynu}
\end{eqnarray}

\subsection{Heavy Majorana Masses}

The leading heavy right-handed neutrino Majorana mass arises from the
operator of Eq.\ref{mop1} where the $\theta $ fields defined in Table \ref%
{Table1} are further Higgs superfields whose vevs break lepton number. It is
spontaneously broken when the right-handed sneutrinos develop vevs in the
third $SU(3)$ direction.This operator gives the Majorana mass, 
\begin{equation}
M_{3}\approx \frac{<\theta >^{2}}{M^{\nu }},  \label{M3}
\end{equation}%
to the third family, where $M^{\nu }=M^{u}$ is the same messenger mass scale
as in the up sector due to $SU(4)_{PS}$. Operators involving $\Sigma $ do
not contribute since it does not couple to right-handed neutrinos which have
zero hypercharge.

The operator in Eq.\ref{mop2} gives Majorana mass, $M_{2}=\epsilon ^{6}\bar{%
\epsilon}^{2}<\theta >^{2}/M,$ to the second family, and the operator in Eq.%
\ref{mop3} gives Majorana mass, $M_{1}=\epsilon ^{6}\bar{\epsilon}%
^{3}<\theta >^{2}/M,$ to the first family, giving the final form 
\begin{equation}
M_{RR}\approx \left( 
\begin{array}{ccr}
\epsilon ^{6}\bar{\epsilon}^{3} & 0 & 0 \\ 
0 & \epsilon ^{6}\bar{\epsilon}^{2} & 0 \\ 
0 & 0 & 1%
\end{array}%
\right) M_{3}.  \label{MRRA}
\end{equation}

\section{Spontaneous Symmetry Breaking\label{spont}}

The pattern of $SU(3)$ family symmetry breaking explored here is as in Eq.%
\ref{fsb}. 
We start with a discussion of the first stage of breaking, $SU(3)\rightarrow
SU(2)$, induced by the vevs of the $\phi_3, \overline{\phi}_3$ Higgs. The
structure of the effective potential is very sensitive to the field content
as well as the additional symmetries. In the context of the model of
interest here the additional symmetry is $U(1)$ or some discrete subgroup of
it which still maintains the structure of the leading operators given in Eqs.%
\ref{op1}-\ref{mop3}. To illustrate the mechanisms that can lead to a
phenomenologically acceptable pattern of symmetry breaking we consider a
simple case in which that discrete subgroup of $U(1)$ is the $Z_{5}\times
Z_3 \times Z_{2}^{\prime}$ symmetry introduced in the last two columns of
Table \ref{Table1}.

As we have discussed it is necessary for there to be a hierarchy in the vevs
of the fields $\phi _{3},\ \overline{\phi _{3}}$ and $\phi _{23},\ \overline{%
\phi }_{23}.$ One way such an hierarchy can develop is through radiative
breaking in which, due to radiative corrections, the running mass squared of
a field becomes negative at some scale$,$ triggering a vev close to this
scale. Gauge interactions increase the mass squared while Yukawa
interactions decrease it so it is likely that the field undergoing radiative
breaking has a reduced gauge symmetry. For this reason we suppose that a ${%
SO(10)\times SU(3)\times Z}_{2R}$ singlet field $S$ acquires a vev due to
(unspecified) Yukawa interactions\footnote{%
It is easy to add a Yukawa interaction involving $X$ and additional fields
to drive radiative breaking. When $X$ acquires a vev these additional field
acquire a large mass and need play no role in low energy phenomenology.}.
Its charge is given in Table \ref{Table2}. We expect the symmetry breaking
will be communicated to the other fields of the theory via heavy messenger
fields. Due to the $Z_{2R}$ symmetry the superpotential does not contain
terms involving the $\phi $ superfields on their own. The only way the
superpotential will generate a potential for these fields is if there are
additional fields carrying $Z_{2R}$ charge $2.$ Allowing for such a ${%
SO(10)\times SU(3)}$ singlet field $U$ carrying the $Z_{5}$ charge as in
Table \ref{Table2} we find the relevant superpotential term $P_{1}$ given by%
\begin{equation}
P_{1}=U((\phi _{23}\overline{\phi }_{23})^{2}+S^{2})  \label{p1}
\end{equation}%
The potential corresponding to the $|F_{U}|^{2}$ term triggers a vev $%
-<S^{2}>\equiv b^{4}$ for the combination $(\phi _{23}\overline{\phi }%
_{23})^{2}$, where $b$ was defined in Eq.\ref{phi23vevs}. Note that in 
Eq.\ref{p1} and in the following equations we have not written the Yukawa
couplings which are expected to be of $O(1)$ and do not change to overall
ordering in terms of the expansion parameter of the mass matrix. However
they are expected to be complex and can generate the phases as in 
Eqs.\ref{phi23vevs},\ref{phibar23vevs} needed for CP violation and the
precise description of the quark masses. Given that the magnitude of the
phases are not determined by the symmetries of the model we do not write
them explicitly here. The vacuum alignment of $\phi _{23}$ will be discussed
below.

In the case of \bigskip $\phi _{3}$ and$\ \overline{\phi _{3}}$ there are
two possibilities:

(i) The first employs the same mechanism as used above, triggering the vevs
by a $SO(10)\times SU(3)\times Z_{2R}$ singlet field, $T$, which also\
acquires a vev through radiative breaking. The superpotential 
\begin{equation}
P_{2}=V(\phi _{3}\overline{\phi }_{3}+T)
\end{equation}
then drives $<\phi _{3}\overline{\phi }_{3}>=-T.$ Finally the relative
magnitudes of the vacuum expectation values of $\phi _{3}$ are fixed by the
following superpotential 
\begin{equation}
P_{3}=WS^{2}\phi _{3}\overline{\phi }_{3}
\end{equation}
which, once the messenger mass scales of the operator are taken into
account, leads to a potential proportional to a sum of squares $%
(a_{3}^{u}/M^{u})^2+(a_{3}^{d}/M^{d})^2$, where the vevs were defined in Eq.%
\ref{phi3vev}. The potential is minimised by $a_{3}^{u}/M^{u}\simeq
a_{3}^{d}/M^{d}$.

(ii) The second possibility applies if there is only a discrete subgroup of
the $U(1)$ in which case, in the absence of the $T$ field, $P_{2}$ takes the
form 
\begin{equation}
P_{2}=V\phi _{3}\overline{\phi }_{3}(1+(\phi _{3}\overline{\phi }%
_{3})^{5}+...)
\end{equation}
The resulting potential has a minimum for $<\phi _{3}\overline{\phi }%
_{3}>=O(1)$ corresponding to the vanishing of the term in brackets which is
a polynomial in $(\phi _{3}\overline{\phi }_{3})^{5}$. In both cases, when
combined with the potential from $P_{3}$ we can generate $<\phi
_{3}^{u}/M^{u}>\simeq $ $<\phi _{3}^{d}/M^{d}>=\sqrt{\overline{\epsilon }}%
=O(1)$, as assumed Eq.\ref{a3}.

{\small 
\begin{table}[tbp] \centering%
\begin{tabular}{|ccccccrr|rrc|}
\hline
${\bf Field}$ & ${\bf SU(3)}$ & ${\bf SU(4)_{PS}}$ & ${\bf SU(2)_{L}}$ & $%
{\bf SU(2)_{R}}$ & ${\bf R}$ & ${\bf Z}_{2}$ & ${\bf U(1)}$ & ${\bf Z}_{5}$
& ${\bf Z}_{3}$ & ${\bf Z}_{2}^{\prime }$ \\ \hline
${\ S}$ & ${\bf 1}$ & ${\bf 1}$ & ${\bf 1}$ & ${\bf 1}$ & ${\bf 0}$ & ${\bf -%
}$ & ${\bf 1}$ & ${\bf 1}$ & ${\bf 1}$ & ${\bf -}$ \\ 
${\ T}$ & ${\bf 1}$ & ${\bf 1}$ & ${\bf 1}$ & ${\bf 1}$ & ${\bf 0}$ & ${\bf +%
}$ & ${\bf -6}$ & ${\bf -1}$ & ${\bf 0}$ & ${\bf +}$ \\ 
${\ U}$ & ${\bf 1}$ & ${\bf 1}$ & ${\bf 1}$ & ${\bf 1}$ & ${\bf 2}$ & ${\bf +%
}$ & ${\bf -2}$ & ${\bf -2}$ & ${\bf 1}$ & ${\bf +}$ \\ 
${\ V}$ & ${\bf 1}$ & ${\bf 1}$ & ${\bf 1}$ & ${\bf 1}$ & ${\bf 2}$ & ${\bf +%
}$ & ${\bf 6}$ & ${\bf 1}$ & ${\bf 1}$ & ${\bf +}$ \\ 
${\ W}$ & ${\bf 1}$ & ${\bf 1}$ & ${\bf 1}$ & ${\bf 1}$ & ${\bf 2}$ & ${\bf +%
}$ & ${\bf 4}$ & ${\bf -1}$ & ${\bf 1}$ & ${\bf +}$ \\ 
${\ X}$ & ${\bf 1}$ & ${\bf 1}$ & ${\bf 1}$ & ${\bf 1}$ & ${\bf 2}$ & ${\bf +%
}$ & -${\bf 1}$ & -${\bf 1}$ & -${\bf 1}$ & ${\bf -}$ \\ 
${\ Y}$ & ${\bf 1}$ & ${\bf 1}$ & ${\bf 1}$ & ${\bf 1}$ & ${\bf 2}$ & ${\bf +%
}$ & ${\bf 7}$ & ${\bf 2}$ & ${\bf 1}$ & ${\bf -}$ \\ 
${\ A}$ & ${\bf 1}$ & ${\bf 1}$ & ${\bf 1}$ & ${\bf 1}$ & ${\bf 2}$ & ${\bf +%
}$ & ${\bf -}$ & ${\bf 1}$ & ${\bf 1}$ & ${\bf +}$ \\ 
${\ \Sigma _{X}}$ & ${\bf 1}$ & ${\bf 15}$ & ${\bf 1}$ & ${\bf 3}$ & ${\bf 2}
$ & ${\bf +}$ & ${\bf -}$ & ${\bf 1}$ & ${\bf 1}$ & ${\bf +}$ \\ 
${\ \theta ^{\prime }}$ & ${\bf \overline{3}}$ & ${\bf 4}$ & ${\bf 1}$ & $%
{\bf 2}$ & ${\bf 2}$ & ${\bf +}$ & ${\bf -}$ & ${\bf -1}$ & ${\bf -1}$ & $%
{\bf +}$ \\ \hline
\end{tabular}%
\caption{{\footnotesize
The charges of the messenger sector fields 
communicating symmetry breaking to the fields in Table 1.
\label{Table2}}} 
\end{table}%
}

There remains the question of the relative vacuum alignment of $\phi _{3}$
and $\phi _{23}.$ This is readily done in the manner proposed in \cite%
{King:2001uz} through the superpotential%
\begin{equation}
P_{4}=X\phi _{3}\overline{\phi }_{2} +\left[ Y(\phi _{23}\overline{\phi _{2}}%
\phi _{23}\overline{\phi _{3}}+(\phi _{3}\overline{\phi }_{23})^{2}(\phi _{3}%
\overline{\phi }_{3})^{2}(\phi _{23}\overline{\phi }_{23})\right]
\end{equation}
where the ${SO(10)\times SU(3)}$ singlet fields $X$, $Y$ \ have the discrete
symmetry charges given in Table \ref{Table2}. The potential from $%
|F_{X}|^{2} $ forces $\phi _{3}$ to be orthogonal to $\overline{\phi }_{2}$
while the potential following from $|F_{Y}|^{2}$ requires that the vev of $%
\phi _{23,2}\phi _{23,3}$ is non-zero. Including soft and D-terms and
minimising the potential then leads to the vacuum alignment of Eqs. \ref%
{phi23vevs},\ref{phibar23vevs} in the manner discussed in \cite{King:2001uz}%
. The alignment is due to the underlying $SU(3)$ symmetry which requires the
soft mass terms of the triplet components to be degenerate.

Finally we consider the $SO(10)$ breaking. As discussed above, gauge
invariant combinations of fields carrying the same discrete quantum numbers
can readily acquire vevs of the same magnitude. Thus, with the introduction
of an additional messenger field, we can readily have $<Tr(\Sigma^{2})>%
\simeq <\phi _{3}\overline{\phi _{3}}>.$ Taking account of the uncertainty
in the messenger mass scale for the $\Sigma $ field this is quite compatible
with the magnitude used in Eq.\ref{sigmad}. Similarly we can readily
construct a messenger sector that drives $<\theta \overline{\theta }>\simeq
<\phi _{3}\overline{\phi _{3}}>.$ Finally the alignment of the $\Sigma $ vev
can be arranged through a term in the potential proportional to $\left\vert
\Sigma \overline{\theta }\right\vert ^{2}$ and again such a term can readily
be generated via a suitable messenger sector. An explicit example (for the
discrete symmetry case) of how this can be done is given by the
superpotential $P_{4}$ with the additional messenger fields given in Table %
\ref{Table2}, 
\begin{equation}
P_{4}=A(\phi _{3}\overline{\phi _{3}}+\theta \overline{\theta }+Tr(\Sigma
^{2}))+\phi _{3}\Sigma _{X}\overline{\phi _{3}}+Tr(\Sigma _{X}\Sigma
^{2})+\theta ^{\prime }\Sigma \overline{\theta }.
\end{equation}

\section{SUSY\ Breaking Soft Terms}

A major problem with a continuous gauged family symmetry is that the D-terms
split the degeneracy needed between the squarks and sleptons. This may give
rise to unacceptably large flavour changing neutral currents (FCNC) due to
flavour dependent contributions to sparticle masses of the form $\Delta 
\tilde{m}_{i}^{2}=c_{i}D^{2}$ where $c_{i}$ is a family dependent
coefficient and $D$ is the magnitude of the D-term. In the model discussed
above the magnitude of the D-terms are small and the FCNC are within
experimental limits. Consider first the D-term which is generated at the
highest scale of symmetry breaking by the fields $\phi _{3}$ and $\overline{%
\phi }_{3}.$ The term in the superpotential $P_{2}$ driving this breaking is
symmetric between the two fields. As a result the magnitude of the $D-$term
obtained from $\overline{\phi }_{3}\partial V/\partial \phi _{3}-\phi
_{3}\partial V/\partial \overline{\phi }_{3}$, where $V$ is the full
potential involving D-terms, scalar mass terms and F-terms, is given by%
\begin{equation}
D_{3}^{2}=\frac{g^{2}}{3}\left\vert \phi _{3}^{\dagger }\phi _{3}-\overline{%
\phi }_{3}^{\dagger }\overline{\phi }_{3}\right\vert ^{2}\simeq \frac{1}{16}%
\left( m_{\overline{3}}^{2}-m_{3}^{2}\right) ^{2}  \label{d3}
\end{equation}%
If the soft masses are driven by supergravity coupling to a hidden
supersymmetry breaking sector and the modular weights of the two fields are
the same, then the masses will be equal at the Planck scale. Since the
breaking triggered by $\phi _{3}$ and $\overline{\phi }_{3}$ is close to the
Planck scale the splitting induced in the soft masses of these fields by
radiative corrections involving Yukawa couplings are likely to be small as
there are no large logarithms involved. As a result the D-term is likely to
be very small.

The second stage of breaking is triggered by the fields $\phi _{23},$ and $%
\overline{\phi }_{23}$. Their D-terms can split the first two generations
and so must be very small if unacceptable flavour changing neutral currents
are to be avoided. In their case the superpotential term, $P_{1},$
triggering their vevs is also symmetric. One one must also include the
effect of the term coming from $P_{3}$ which spoils the symmetry between $%
\phi _{3}$ and $\overline{\phi }_{3}$ and between $\phi _{23},$ and $%
\overline{\phi }_{23}.$ Together this gives for the $D-term$ associated with
the second stage of symmetry breaking the form 
\begin{equation}
D_{23}=\frac{1}{4}\left( m_{\overline{23}}^{2}-m_{23}^{2}\right) -2\overline{%
\phi }_{2}\overline{\phi }_{3}\phi _{23}\phi _{23}^{\dagger }\left( 
\overline{\phi }_{2}\overline{\phi }_{3}\phi _{23}\phi _{23}-\mu ^{4}\right)
^{\dagger }
\end{equation}%
Since the vev of $F_{Y}=\left( \overline{\phi }_{2}\overline{\phi }_{3}\phi
_{23}\phi _{23}-\mu ^{4}\right) $ is of order the soft mass squared the
second term is small, being suppressed by the small vevs of $\overline{\phi }%
_{2}$ and $\phi _{23}.$ The first term is also expected to be small if the
soft masses of $\phi _{23}$ and $\overline{\phi }_{23}$ are degenerate at
the Planck scale because $\phi _{23}$ and $\overline{\phi }_{23}$ also have
vevs close to the Planck scale and so the radiative corrections splitting
these masses are likely to be small.

The general property that keeps D-terms small is the fact that to a good
approximation the potential is symmetric in the conjugate fields. This
immediately leads to the form given by Eq.(\ref{d3}). For breaking close to
the Planck scale radiative corrections splitting the soft masses will be
suppressed by the one loop expansion parameter and may readily be small. In
this case any supersymmetry breaking mechanism giving degenerate soft SUSY
breaking masses at the Planck scale to the conjugate fields involved in the
family symmetry breaking leads to small D-terms consistent with the bounds
from FCNC. This mechanism can be applied more generally to family symmetry
models and removes one of the major obstacles to implementing such a
symmetry.

\section{Phenomenology}

The model we have constructed gives excellent agreement with the quark and
lepton masses and mixing angles. For the up and down quarks the form of $%
Y^{u}$ and $Y^{d}$ given in Eq.\ref{Yu}, \ref{Yd} is consistent with the
phenomenological fit in Eq.\ref{yuk}, with the expansion parameters as in Eq.%
\ref{exp}, for parameters such as $g\sim 1$, $g'\sim -1$, 
$h\sim h'\sim 0.3$, $c^{\prime }\sim 1$, $c\sim -1$. 
The charged lepton mass matrix is of the Georgi Jarslkog form which, after
including radiative corrections, gives an excellent description of the
charged lepton masses. In the neutrino sector the parameters satisfy the
conditions of sequential dominance, with the lightest right-handed neutrino
giving the dominant contribution to the heaviest physical neutrino mass, and
the second right-handed neutrino giving the leading subdominant
contribution, providing that $\alpha \sim \epsilon $. 

Analytic estimates of neutrino masses and mixing angles for sequential
dominance were derived in \cite{King:2002nf}, and for the special case here
of light sequential dominance, with the 11 neutrino Yukawa coupling equal to
zero, they are summarized recently in \cite{King:2002qh}, from which we
readily extract the analytic estimates below for the neutrino masses, 
\begin{eqnarray}
m_{1} &\sim &\bar{\epsilon}^{2}\frac{v_{2}^{2}}{M_{3}} \\
m_{2} &\approx &\frac{(g+\frac{h}{3}+\frac{h'}{3})^{2}}{s_{12}^{2}}\frac{v_{2}^{2}}{M_{3}}%
\sim 5.8\frac{v_{2}^{2}}{M_{3}} \\
m_{3} &\approx &\frac{[(g'-\frac{h}{3}-\frac{h'}{3})^{2}
+(g'+\frac{h}{3})^{2}]}{\bar{%
\epsilon}}\frac{v_{2}^{2}}{M_{3}}\sim 15\frac{v_{2}^{2}}{M_{3}}
\end{eqnarray}%
and neutrino mixing angles: 
\begin{eqnarray}
\tan \theta _{23}^{\nu } &\approx &\frac{(g'-\frac{h}{3}-\frac{h'}{3})}{(g'+\frac{h}{3})}%
\sim 1.3 \\
\tan \theta _{12}^{\nu } &\approx &(\bar{\epsilon})^{1/2}
\frac{(g+\frac{h}{3}+\frac{h'}{3})}{-cs_{23}}\sim 0.66 \\
\theta _{13}^{\nu } &\approx &-(\bar{\epsilon})
\frac{(g+\frac{h}{3}+\frac{h'}{3})[(g'-\frac{h}{3}-\frac{h'}{3})
(-\frac{\alpha }{\epsilon })
+(g'+\frac{h}{3})(-\frac{\alpha }{\epsilon }+c\bar{\epsilon}^{-1/2})]}
{[(g'-\frac{h}{3}-\frac{h'}{3})^{2}+(g'+\frac{h}{3})^{2}]^{3/2}}
\sim 1.6\bar{\epsilon}
\end{eqnarray}%
where the numerical estimates correspond to $g\sim 1$, $g'\sim -1$, 
$h\sim h'\sim 0.3$, $c^{\prime }\sim -c\sim 1$, 
$\alpha \sim \epsilon \sim 0.05$, $\bar{\epsilon}\sim 0.15$. 
Note that the physical lepton mixing angle $\theta_{13}$
receives a large contribution from the neutrino sector $\theta_{13}^{\nu}%
\sim 0.3$ at the high energy scale, for this choice of parameters, compared
to the current CHOOZ limit $\theta_{13}\leq 0.2$ \cite{Apollonio:1999ae}.
However the physical mixing angles will receive charged lepton contributions 
\cite{King:2002nf} and all the parameters are subject to radiative
corrections in running from the high energy scale to low energies, although
in sequential dominance models these corrections are only a few per cent 
\cite{King:2000hk}. We conclude that our model predicts that $\theta_{13}$
is close to the current CHOOZ limit, and could be observed by the next
generation of long baseline experiments such as MINOS or OPERA.

Any model of flavour must be sure to avoid large FCNC. In the previous
Section we showed that the D-terms were under control in this respect.
However there is also a problem in supergravity models which use the
Froggatt Nielsen mechanism to order fermion masses due to the fact that the
Froggatt Nielsen fields typically acquire a F-term vevs which cause the $A$
terms to be misaligned by $O(m_{3/2})$ relative to the Yukawa couplings,
generating FCNC and potentially large electric dipole moments \cite%
{Ross:2002mr} (see also \cite{Abel:2001cv}). 
The effect is somewhat ameliorated here because the mass
matrices are symmetric. However we note that lepton number violating
processes are still expected to occur at a rate close to the present limits.

In any scheme such as this in which the neutrino Dirac mass is equal to the
up Dirac mass and the dominant right handed neutrino exchange is in the $1$
direction we have a prediction for the lightest Majorana state given by $%
M_{1}=m_{u}m_{c}/m_{3}\simeq 10^{8}\ {\rm GeV}$. In this model the expansion
parameter is $\propto \varepsilon ^{6}$ so a very small difference in the
neutrino exapansion parameter, coming from $SU(4)$ breaking, can readily
increase this by more than a factor of 10, bringing it into the range that
thermal leptogenesis is possible. Assuming that this is the case the
resultant CP asymmetry in the decay of the heavy lepton may be readily
estimated giving 
\begin{eqnarray*}
\epsilon _{1} &\simeq &-\frac{3}{8\pi ^{2}}{Im}\left( (YY^{\dagger
})_{11}\right) \\
&\simeq &-\frac{3}{8\pi ^{2}}\epsilon ^{6}
\end{eqnarray*}%
This gives the asymmetry $\epsilon _{1}=6.10^{-10}$ or somewhat larger if we
allow for a slightly larger expansion parameter in the neutrino sector.
However washout effects can significantly reduce the asymmetry. An analysis
of these effects in this class of model is given in \cite{ibarra}. Note that
in such models there is a link between the leptogenesis CP violating phase
and the neutrino mixing phase measurable in neutrino oscillation experiments 
\cite{King:2002qh}.

\section{Summary and Conclusions}

The main message of this paper is that a coherent description of all quark,
charged lepton and neutrino masses and mixing angles can be constructed in a
model having a very high degree of symmetry. The large mixing angles in the
lepton sector follow naturally from two ingredients, the see-saw mechanism
with sequential right handed neutrino dominance and a non-Abelian family
symmetry. We have constructed a simple implementation of these ideas in
which there is an underlying stage of $SO(10)$ type Unification and the
Family Symmetry is $SU(3)$ together with a further discrete symmetry needed
to restrict the allowed Yukawa couplings. The resultant model gives an
excellent description of all data including lepton mixing consistent with
almost maximal atmospheric mixing and the solar LMA MSW solution. The model
naturally satisfies the bounds on flavour changing neutral currents. Indeed
the $SU(3)$ symmetry provides a new mechanism for making the families of
squarks and sleptons of a given flavour degenerate. Note that this does not
require that all squarks and sleptons (and Higgs) be degenerate as in the
SUGRA solution of the flavour problem. Indeed it provides a new solution to
the flavour problem with different expectations for the SUSY\ spectrum from
SUGRA, gauge and anomaly mediated schemes. In particular the strong breaking
of $SU(3)$ in the third family direction, needed to give the large top and
bottom quark masses, will give large splitting to the third generation
through the terms $\left\vert \psi _{i}\phi _{3}^{i}\right\vert ^{2},$ $%
\left\vert \psi _{i}^{c}\phi _{3}^{i}\right\vert \cite{King:2001uz}.$ This
can considerably affect the allowed region of parameter space in a
constrained model fit and the spectrum of supersymmetric states 
\cite{Ramage:2003pf}. 
As we discussed above there is still the need for the fields $\phi _{3}$
and $\overline{\phi }_{3}$ and the fields $\phi _{23}$ and $\overline{\phi }%
_{23}$ to have equal initial masses respectively, but even this condition
can be relaxed if one modifies the model to use a discrete subgroup of $%
SU(3) $ instead of the full family group, because then the D-terms
associated with the family symmetry are absent.

\begin{center}
{\bf Acknowledgement}
\end{center}

One of us (GGR) would like to thank A.Ibarra, L. Velasco-Sevilla, S.
Pokorski, R. Rattazzi, P. Ramond and particularly O. Vives for helpful
discussions. SFK thanks T. Blazek for discussions, I. Peddie for carefully
reading the manuscript and correcting some errors,
and is grateful to PPARC for the support of a Senior
Fellowship. This work was partly supported by the EU network, "Physics
Across the Present Energy Frontier HPRV-CT-2000-00148.

\bigskip

\end{document}